\documentclass[showkeys,showpacs,nofootinbib,superscriptaddress]{revtex4-2}
\usepackage{amsmath}
\usepackage{amssymb}
\usepackage{graphicx}
\usepackage[utf8]{inputenc}
\usepackage[russian]{babel}
\usepackage{float}
\usepackage{xcolor}
\usepackage{physics}
\usepackage{ulem}

\allowdisplaybreaks

\begin{document}

\title{
 Chromoelectric flux tubes within non-Abelian Proca theory
}

\author{
Vladimir Dzhunushaliev
}
\email{v.dzhunushaliev@gmail.com}
\affiliation{
Department of Theoretical and Nuclear Physics,  Al-Farabi Kazakh National University, Almaty 050040, Kazakhstan
}
\affiliation{
Academician J.~Jeenbaev Institute of Physics of the NAS of the Kyrgyz Republic, 265 a, Chui Street, Bishkek 720071, Kyrgyzstan
}

\author{Vladimir Folomeev}
\email{vfolomeev@mail.ru}
\affiliation{
Academician J.~Jeenbaev Institute of Physics of the NAS of the Kyrgyz Republic, 265 a, Chui Street, Bishkek 720071, Kyrgyzstan
}
\affiliation{
International Laboratory for Theoretical Cosmology, Tomsk State University of Control Systems and Radioelectronics (TUSUR),
Tomsk 634050, Russia
}


\begin{abstract}
Flux tube solutions within non-Abelian SU(3) Proca theory with external sources are obtained. It is shown that such tubes have a longitudinal chromoelectric field possessing two components (nonlinear and gradient), as well as a transverse chromomagnetic field whose force lines create concentric circles with the center on the axis of the tube. The scenario of a possible relationship between non-Abelian Proca theory and quantum 
chromodynamics is considered. In such scenario: 
(a)~the components of color fields have different behavior:  those which are almost classical, and those which are purely quantum;
(b)~the second components create a gluon condensate that is a source of the field for the almost classical components of the Proca field; 
(c)~Proca masses may appear as a result of an approximate description of the gluon condensate; 
(d)~the question of gauge invariance is considered. 
It is shown that the results obtained are in good agreement with the results of lattice calculations. We make an assumption that an approximate description of a flux tube in quantum chromodynamics can be carried out using classical Proca equations but with a mandatory account of a gluon condensate. 
\end{abstract}

\pacs{11.90.+t
}

\keywords{
	non-Abelian Proca theory, external sources, flux tube, total energy, linear dependence
}
\date{\today}

\maketitle

\section{Introduction}

One of manifestations of the confinement phenomenon in quantum chromodynamics (QCD) is the presence of a tube filled with a longitudinal chromoelectric field created by quarks located at the ends of the tube. This phenomenon has been supported by lattice calculations (see, e.g.,  Refs.~\cite{DiGiacomo:1990hc,Bali:1994de,Simonov:1996ati,Cea:2012qw,Cardoso:2013lla,Baker:2018mhw,Baker:2021jnr} 
and references therein). In particular, in those works, it was shown that between static quarks there appears a flux tube threaded by chromoelectric and chromomagnetic fields, whose structure and a relationhip between the fields and currents were studied. These calculations are based on the direct application of numerical methods for nonperturbative quantization, i.e., for the calculation of the corresponding path integral. 
However, in order to clearly appreciate physical processes, one would like to have some approximate model describing the appearance of the flux tube between quarks filled with a longitudinal chromoelectric field.

In the present paper we show that in non-Abelian Yang-Mills-Proca theory there exist such solutions describing flux tubes supported by external sources. In recent years,  there is a growing interest in Proca fields inspired by the possibility of obtaining within such theories new solutions that can be used for a 
description of various physical objects and processes.  For example, the literature in the field suggests gravitating compact 
configurations~\cite{Brito:2015pxa,Herdeiro:2017fhv,Dzhunushaliev:2019kiy,Herdeiro:2019mbz,Dzhunushaliev:2019uft,Bustillo:2020syj,Dzhunushaliev:2021vwn}, various aspects of black hole physics~\cite{Heisenberg:2017xda,Kase:2018voo}, dark matter models~\cite{Arkani-Hamed:2008hhe,Pospelov:2008jd}, a study of processes at cosmological scales~\cite{DeFelice:2016yws,DeFelice:2016uil}, a consideration of effects arising due to the possible presence of the rest mass of a photon~\cite{Tu:2005ge}, generalized Proca theories with second-order equations of motion in a curved spacetime \cite{DeFelice:2016cri}. Also, as applied to modelling flux tubes, Proca fields have been used by us earlier in constructing various tubelike configurations filled with electric and magnetic  fields~\cite{Dzhunushaliev:2019sxk,Dzhunushaliev:2020eqa,Dzhunushaliev:2021uit,Dzhunushaliev:2021oad,Dzhunushaliev:2021tlm}.

A distinctive feature of the configurations considered in the present work is that they contain a flux of chromoelectric field directed from one static source to another. This field consists of two components, one of which is nonlinear, and therefore it may be called nonperturbative, while another one is gradient, and it is similar to an electric field in Maxwell's electrodynamics. Another feature of the systems obtained here is that the total energy of such field configurations increases linearly with increasing distance between the static sources.

These properties of the solutions enable us to assume that there can be some relationship between non-Abelian Proca theory under consideration and QCD. In particular, one can assume that non-Abelian Proca theory is some approximation for nonperturbative QCD. Such an assumption leads to interesting consequences (for details see the discussion in Sec.~\ref{discussion}): 
(a)~the components of color fields have different behavior:  those which are almost classical (with a nonzero quantum average), and those which are purely quantum (with a zero quantum average);
(b)~the second components create a gluon condensate that is a source of the field for the almost classical components of the Proca field; 
(c)~Proca masses may appear as a result of an approximate description of the gluon condensate.
 
The paper is organized as follows. In Secs.~\ref{Proca_theory} and \ref{gen_eqs_Proca}, we write down the general field equations for non-Abelian Proca theory employed. In Sec.~\ref{ft_sols}, we solve these equations numerically and obtain axially symmetric regular solutions describing localized flux tube configurations consisting of chromoelectric and chromomagnetic Proca fields sourced by external charges and currents. In Sec.~\ref{comparison}, we compare our computations with the results of lattice calculations. In Sec.~\ref{discussion}, we discuss a possible scenario describing the relationship between 
non-Abelian Proca theory under consideration and QCD. Finally, Sec.~\ref{concl}  summarizes the results obtained in the paper.

\section{Non-Abelian SU(3) Yang-Mills-Proca theory}
\label{Proca_theory}

The Lagrangian describing a system consisting of a non-Abelian SU(3) Proca field $A^a_\mu$
sourced by color charges and currents can be taken in the form (hereafter, we work in natural units $c=\hbar=1$)
\begin{equation}
	\mathcal L =  - \frac{1}{4} F^a_{\mu \nu} F^{a \mu \nu} +
	\frac{m^2}{2}	A^a_\mu A^{a \mu} - A^a_\mu j^{a \mu} .
\label{0_10}
\end{equation}
Here
$
	F^a_{\mu \nu} = \partial_\mu A^a_\nu - \partial_\nu A^a_\mu +
	g f_{a b c} A^b_\mu A^c_\nu
$ is the field strength tensor for the Proca field of mass $m$, 
where $f_{a b c}$ are the SU(3) structure constants, $g$ is the coupling constant, $j^{a \mu}$ are color currents,
 $a,b,c = 1,2, \dots, 8$ are color indices, $\mu, \nu = 0, 1, 2, 3$ are spacetime indices. 

Using \eqref{0_10}, the corresponding field equations can be written in the form:
\begin{equation}
D_\nu F^{a \mu \nu} - 
			 m^2 A^{a \mu}\equiv
\frac{1}{\sqrt{-\mathcal G}} \frac{\partial}{\partial x^\nu}\left(\sqrt{-\mathcal G}F^{a \mu \nu}\right) + g f_{a b c} A^b_\nu F^{c \mu \nu}
	- m^2 A^{a \mu} = - j^{a \mu},
\label{0_20}
\end{equation}
where $\mathcal G$ is the determinant of the spacetime metric,
and the energy density is
\begin{equation}
\begin{split}
	\varepsilon = &\frac{1}{2} \left( E^a_i \right)^2 +
	\frac{1}{2} \left( H^a_i \right)^2 +
	m^2\left( 
		 A^{a 0} A^a_0 -
		\frac{1}{2}  A^a_\alpha A^{a \alpha}
	\right) ,
\end{split}
\label{energy_dens}
\end{equation}
where $i=1,2,3$ and $E^a_i$ and $H^a_i$ are the components of the color electric and magnetic field strengths, respectively.

\section{General equations and Ansatz
}
\label{gen_eqs_Proca}

In this section we write out the field equations within the theory  \eqref{0_10} in some generalized form in order to analyze the possibility of obtaining 
flux tube solutions. Namely, to find a distribution of the non-Abelian Proca field created by static charges, one can choose the following Ansatz
in cylindrical coordinates $(t, \rho, \varphi, z )$: 
\begin{equation}
	A^2_t = \frac{f(\rho, z)}{g} , \;
	A^2_\varphi = \frac{\rho k(\rho, z)}{g} , \;
	A^5_\rho = \frac{u(\rho, z)}{g}, 
	A^5_z = \frac{v(\rho, z)}{g}, 
	A^7_t = \frac{h(\rho, z)}{g} , \; 
	A^7_\varphi = \frac{\rho w(\rho, z)}{g} . 
\label{1_10}
\end{equation}
Henceforth  the upper indices $2, 5, 7$ are color indices, and one can introduce the vector $\vec{A}^5 = \frac{1}{g} \left\{ u(\rho, z), 0, v(\rho, z) \right\} $. 
The potentials \eqref{1_10} give the following nonzero components of color electric and magnetic field strengths: 
\begin{align}
	E^2_z = & -\frac{h v}{2 g} - \frac{f_{, z}}{g} , \quad
	E^7_z = \frac{f v}{2 g} -\frac{h_{, z}}{g}, \quad
	E^2_\rho = - \frac{h u}{2 g} - \frac{f_{, \rho}}{g} , \quad
	E^7_\rho = \frac{f u}{2 g} -\frac{h_{, \rho}}{g}, \quad
	E^5_\varphi = \frac{\rho}{2 g} 
	\left( h k	- f w \right) ,
\label{fields_5}\\
	H^2_z = & - \frac{
		2 \left( k_{, \rho} + \frac{k}{\rho}\right)  + u w
	}{2 g} , \quad
	H^7_z = - \frac{
		2 \left(w_{, \rho} + \frac{w}{\rho}\right) - u k
	}{2 g} , \quad
	H^2_\rho =  \frac{2 k_{, z}  + v w}{2 g} ,  \quad
\nonumber \\
	H^7_\rho = & \frac{\left(2 w_{, z} - k v\right)}{2 g} ,  \quad
	H^5_\varphi = \rho 
	\frac{ \left(v_{, \rho} - u_{, z}\right)}{g} , 
\label{fields_10}
\end{align}
where a comma in lower indices denotes differentiation with respect to the corresponding coordinate.

The Poynting vector created by crossed color electric and magnetic fields is
$$
	S^i = \frac{\epsilon^{i j k}}{\sqrt{\gamma}} E^a_j H^a_k ,
$$
where  $\gamma$ is the determinant of the space metric and $\epsilon^{i j k}$ is the completely antisymmetric Levi-Civita symbol. 
For the field strengths~\eqref{fields_5} and \eqref{fields_10}, this expression yields the following nonvanishing component:
\begin{equation}
\begin{split}
	g^2 \rho S_\varphi = & k \left[ 
		- \frac{f_{, \rho }}{\rho } - \frac{f}{4}\left( u^2 + v^2 \right) + 
		\frac{u}{2} \left(
			h_{, \rho } - \frac{h}{\rho }
			\right)
			+ \frac{v h_{, z}}{2}
	\right] 
	-	k_{, \rho } \left(f_{, \rho } + \frac{h u}{2}\right)
	- k_{, z} \left( f_{, z} + \frac{h v}{2}\right)
\\
	+ & w \left[ 
		- \frac{h_{, \rho }}{\rho } - \frac{h}{4} \left( u^2 + v^2 \right) 
		- \frac{u}{2} \left( f_{, \rho } - \frac{f}{\rho }\right)
		- \frac{v f_{, z}}{2} 
	\right] 
	- w_{, \rho } \left(h_{, \rho } - \frac{f u}{2} \right)
	- w_{, z} \left(h_{, z} - \frac{f v}{2}\right) .
\end{split}
\label{UmPoynt}
\end{equation}

The Proca equations  \eqref{0_20} for the Ansatz \eqref{1_10} have the form:
\begin{align}
	f_{, zz} + f_{, \rho \rho } + \frac{f_{, \rho }}{\rho } 
	- \frac{f}{4} \left(
		 u^2 + v^2 + w^2 
	\right) 
	+ \frac{h k w}{4}
	+ u h_{, \rho } + \frac{h u_{, \rho }}{2} 
	+ v h_{, z} + \frac{h v_{, z}}{2} + \frac{h u}{2 \rho }	
	- m^2 f = & - g j^2_t , 
\label{1_20}\\
	k_{, zz} + k_{, \rho \rho } + \frac{k_{, \rho }}{\rho } 
	- \frac{k}{\rho ^2} + 
	\frac{k}{4} \left(
		h^2 - u^2 - v^2 
	\right) 
	- \frac{1}{4} f h w 
	+ \frac{w u_{, \rho }}{2} + u w_{, \rho } 
	+ \frac{w v_{, z}}{2} + v w_{, z} 
	+ \frac{u w}{2 \rho } 
	- m^2 k 
	=  & g \rho j^2_\varphi , 
\label{1_30}\\
	v_{, \rho \rho } + \frac{v_{, \rho }}{\rho } 
	- u_{, z\rho } - \frac{u_{, z}}{\rho } 
	+ \frac{v}{4} \left( 
		f^2 + h^2 - k^2 - w^2 
	\right) 
	+ \frac{h f_{, z}}{2} - \frac{f h_{, z}}{2} +\frac{k w_{, z}}{2}  -\frac{w k_{, z}}{2} 
	- m^2 v = &  g j^5_v , 
\label{1_40}\\
	u_{, zz} - v_{, z \rho} 
	+ \frac{u}{4} \left( 
		f^2 + h^2 - k^2 - w^2 
	\right)	
	+ \frac{h f_{, \rho }}{2} - \frac{f h_{, \rho }}{2} +\frac{k w_{, \rho}}{2} -\frac{w k_{, \rho}}{2} 
	- m^2 u 
	= & g j^5_u , 
\label{1_50}\\
	h_{, zz} + h_{, \rho \rho } + \frac{h_{, \rho }}{\rho } 
	- \frac{h}{4} \left(
		u^2 + v^2 + k^2 
	\right) 
	+ \frac{f k w}{4} 
	- u f_{, \rho } - \frac{f u_{, \rho }}{2} - v f_{, z} - \frac{f v_{, z}}{2} 
	- \frac{f u}{2 \rho } 
	- m^2 h = & - g j^7_t , 
\label{1_60}\\
  w_{, zz} + w_{, \rho \rho } + \frac{w_{, \rho }}{\rho } 
  - \frac{w}{\rho^2}
  + \frac{w}{4} \left(
  	f^2 - u^2 - v^2 
  \right) 
  - \frac{1}{4} f h k 
  - \frac{k u_{, \rho }}{2} - u k_{, \rho } - \frac{k v_{, z}}{2} - v k_{, z} 
  - \frac{k u}{2 \rho } 
  - m^2 w 
	= & g \rho j^7_\varphi . 
\label{1_70}
\end{align}

Let us calculate the covariant divergence of the Proca equations~\eqref{0_20},
\begin{equation}
	D_\mu \left( 
		D_\nu F^{a \mu \nu} - 
			 m^2 A^{a \mu}
	\right) = - 
	D_\mu j^{a \mu} . 
\label{1_100}
\end{equation}
Taking into account Eqs.~\eqref{1_20}-\eqref{1_70}, we then have from Eq.~\eqref{1_100}: 
\begin{equation}
	m^2 \div \vec{A}^5 \equiv m^2 \left( v_{, z} + u_{, \rho } + \frac{u}{\rho} \right) 
	= \frac{1}{2} \left(f j^7_t -  h j^2_t\right) 
	+ \frac{1}{2} \rho \left(k j^7_\varphi - w j^2_\varphi\right)
	- \left( j^5_u\right)_{, \rho} - \frac{j^5_u}{\rho } - \left( j^5_v\right)_{,z}  .
\label{1_120}
\end{equation}

The conservation of the four-current $j^{a\mu}$ implies that the components of the current must be chosen so as to make the right-hand side of Eq.~\eqref{1_120} be identically equal to zero. 
Then, using Eq.~\eqref{1_120}, one can get rid of mixed derivatives in Eqs.~\eqref{1_40} and \eqref{1_50}. As a result, the full system of equations \eqref{1_20}-\eqref{1_70} 
for the functions $f, h, u, v, k,$ and $w$  [on account of the divergent equation~\eqref{1_120}] takes the form:
\begin{align}
	\laplacian{f}
	- \frac{f}{4} \left(
		 u^2 + v^2 + w^2 
	\right) 
	+ \frac{h k w}{4}
	+ \vec{A}^5 \cdot \grad{h} - m^2 f 
	= & -g j^2_t ,
\label{1_150}\\
	\laplacian{k} 
	- \frac{k}{\rho ^2} + 
	\frac{k}{4} \left(
		h^2 - u^2 - v^2 
	\right) 
	- \frac{1}{4} f h w 
	+ \vec{A}^5 \cdot \grad{w} 
	- m^2 k 
	= & g\rho j^2_\varphi , 
\label{1_165}\\
	\laplacian{v} 
	+ \frac{v}{4} \left( 
		f^2 + h^2 - k^2 - w^2 
	\right) 
	+ \frac{h f_{, z}}{2} - \frac{f h_{, z}}{2}  
	+ \frac{k w_{, z}}{2} - \frac{w k_{, z}}{2} 
	- m^2 v 
	= & g j^5_v , 
\label{1_160}\\
	\laplacian{u} - \frac{u}{\rho^2} 
	+ \frac{u}{4} \left( 
		f^2 + h^2 - k^2 - w^2  
	\right) 
	+ \frac{h f_{, \rho }}{2} - \frac{f h_{, \rho }}{2} 
	+ \frac{k w_{, \rho }}{2} - \frac{w k_{, \rho }}{2} 
	- m^2 u 
	= & g j^5_u , 
\label{1_170}\\
	\laplacian{h} 
	- \frac{h}{4} \left(
		u^2 + v^2 + k^2 
	\right) 
	+ \frac{f k w}{4} 
	- \vec{A}^5 \cdot \grad{f} - m^2 h 
	= & -g j^7_t , 
\label{1_180}\\
  \laplacian{w} 
  - \frac{w}{\rho^2}
  + \frac{w}{4} \left(
  	f^2 - u^2 - v^2 
  \right) 
   - \frac{1}{4} f h k 
  - \vec{A}^5 \cdot \grad{k}
  - m^2 w 
	= & g\rho j^7_\varphi , 
\label{1_185}
\end{align}
where $\laplacian = \partial_{zz} + \partial_{\rho \rho} + \partial_\rho/\rho$ is the Laplacian in cylindrical coordinates. 

\section{Flux tube solutions}
\label{ft_sols}

To find a flux tube connecting static charges, consider a simplified system  where in Eqs.~\eqref{1_150}-\eqref{1_185} the vector potentials
$w=k=0$. As a result, there will be the following reduced set of equations for the functions $f, h, v,$ and $u$:
\begin{align}
	\laplacian{f}
	- \frac{f}{4} \left(
		u^2 + v^2 
	\right) 
	+ \vec{A}^5 \cdot \grad{h} -  f 
	= & - j^2_t ,
\label{eqn_f}\\
	\laplacian{v} 
	+ \frac{v}{4} \left( 
		f^2 + h^2 
	\right) 
	+ \frac{h f_{, z}}{2} - \frac{f h_{, z}}{2} 
	-  v 
	= &  j^5_v , 
\label{eqn_v}\\
	\laplacian{u} - \frac{u}{\rho^2} 
	+ \frac{u}{4} \left( 
		f^2 + h^2 
	\right) 
	+ \frac{h f_{, \rho }}{2} - \frac{f h_{, \rho }}{2} 
	-  u 
	= &  j^5_u , 
\label{eqn_u}\\
	\laplacian{h} 
	- \frac{h}{4} \left(
		u^2 + v^2 
	\right) 
	- \vec{A}^5 \cdot \grad{f} -  h 
	= & - j^7_t , 
\label{eqn_h}
\end{align}
and in this case the Poynting vector \eqref{UmPoynt} is equal to zero.

The equations~\eqref{eqn_f}-\eqref{eqn_h} are written already in terms of the dimensionless variables
$$
	\left( \rho, z\right) \to \frac{1}{m} \left( \rho, z\right) , \quad
	\left(  f,   h,   u,   v \right) \to m \left(f,  h, u, v \right) , \quad
    \left(  j^{2}_t,   j^{7}_t, j^{5}_u ,  j^{5}_v\right) \to \frac{m^3}{g} \left( j^{2 }_t,   j^{7}_t,  j^{5 }_u , j^{5 }_v \right) .
$$

For our purpose, nonzero components of the four-current are chosen by giving the following {\it ad hoc} expressions:
\begin{align}
	j^2_t = & -\exp \left(- \frac{\rho^2}{\rho_1^2}\right) \left\lbrace 
		j_1 \exp \left[
		- \frac{(z - z_0)^2}{z_1^2} 
	\right] 
		+ j_2 \exp \left[
			- \frac{(z + z_0)^2}{z_1^2} 
		\right]
	\right. 
\nonumber\\
	& \left. +	j_{\text{glc}}
	\left[ 
		\frac{1}{1+\exp\left[-10(z+z_0)\right]}+\frac{1}{1+\exp\left[-10(z-z_0)\right]}-1
	\right] 
	\right\rbrace , 
\label{current_2t}\\ 
	j^5_u = &  \rho  \left\lbrace 
		j_3 \exp \left[ - \frac{\rho^2}{\rho_1^2} - \frac{(z - z_0)^2}{z_1^2}
		\right] 
		+ j_4 \exp \left[ - \frac{\rho^2}{\rho_1^2} - \frac{(z + z_0)^2}{z_1^2}
		\right] 
	\right\rbrace .
\label{current_5_rho}
\end{align}
Here $j_{1, 2, 3, 4}$, $\rho_{1}$, and $z_{1}$ are some parameters characterizing linear sizes of color charge and current densities;
 $z_0$~characterizes the distance between static charges; $j_{\text{glc}}$ is a constant determining the magnitude of a ``gluon condensate''  (cf. Sec.~\ref{discussion}).
 The latter is modeled in the form of some distribution of constant charge density which fills the space between the static charges. 
 
The law of conservation of the current $D_\mu j^{a \mu} = 0$ following from Eq.~\eqref{1_100} implies that in our case Eq.~\eqref{1_120} must hold. 
In our model problem we solve this equation as follows:
	\begin{align}
		v_{, z} + u_{, \rho } + \frac{u}{\rho} = & 0 , 
		\label{div_eqn_1}\\
		f j^7_t - h j^2_t =& 0, 
		\label{div_eqn_2}\\
		\left( j^5_u\right)_{, \rho} + \frac{j^5_u}{\rho } + \left( j^5_v\right)_{,z} 
		= & 0 . 
		\label{div_eqn_3}
	\end{align}
That is, we take  $\div \vec{A}^5 = 0$, since for the currents  $j^{a \mu}$ created by fields (spinor or scalar) the equation $D_\mu j^{a \mu} = 0$ must hold.
Then, taking into account the expressions \eqref{current_2t} and \eqref{current_5_rho}, we have from
Eqs.~\eqref{div_eqn_2} and \eqref{div_eqn_3} the following expressions for the components of the four-current $j^7_t$ and $j^5_v$:
\begin{align}
	j^7_t = & \frac{h}{f} j^2_t , 
\label{current_7t}\\ 
	j^5_v = &  \frac{\sqrt{\pi } z_1 }{\rho_1^2}
	\exp (- \frac{\rho ^2}{\rho_1^2}) 
	\left(\rho^2 - \rho_1^2\right) 
	\left[ 
		j_3 \erf \left(\frac{z - z_0}{z_1}\right) 
		+ j_4 \erf \left(\frac{z + z_0}{z_1}\right)
	\right] . 
\label{current_5z}
\end{align}

\subsection{Boundary conditions and numerical approach
}
The set of nonlinear partial differential equations \eqref{eqn_f}-\eqref{eqn_h} is solved numerically by specifying  
appropriate boundary conditions on the region of integration $0\leq \rho <\infty$, $-\infty < z < \infty$. 
In order to have a color longitudinal electric field directed along the axis of the tube, we choose the  boundary conditions as follows: 
\begin{equation}
\begin{split}
	&\left. \frac{\partial f}{\partial \rho}\right|_{\rho = 0} =
	\left. \frac{\partial h}{\partial \rho}\right|_{\rho = 0} =\left. \frac{\partial v}{\partial \rho}\right|_{\rho = 0}=  0,\quad  \left. u \right|_{\rho = 0} = 0;
	 \\
	& f = h = u = v = 0  \quad \text{ as } \quad \rho^2 +z^2 \to \infty . \nonumber
\end{split}
\end{equation}
These boundary conditions enable us to obtain solutions regular over the whole space. 

For numerical calculations, it is convenient to introduce new compactified coordinates
\begin{equation}
	\bar \rho=\frac{2}{\pi}\arctan \rho,\quad \bar z=\frac{2}{\pi}\arctan z,
	\label{comp_coord}
\end{equation}
the use of which permits one to map the infinite region $(-\infty,\infty)$ to the finite interval $[-1,1]$.

All results of numerical calculations given below have been obtained using 
the Intel MKL PARDISO sparse direct solver \cite{pardiso1,pardiso2} and the CESDSOL library\footnote{Complex Equations-Simple Domain partial differential equations SOLver, a C++ package developed by I.~Perapechka,
see, e.g., Ref.~\cite{Kunz:2019sgn}.}.
Solutions are sought on a grid of $200\times 400$ points, covering  the region of integration $0\leq \bar \rho \leq 1$ and $-1\leq \bar z \leq 1$
given by the compactified coordinates from Eq.~\eqref{comp_coord}. This enables us to get solutions with typical relative errors on the order of $10^{-4}$.
Also,  in performing calculations, we have been checking the fulfilment of  the divergent equation~\eqref{div_eqn_1}.

\begin{figure}[t]
	\begin{center}
		\includegraphics[width=1.\linewidth]{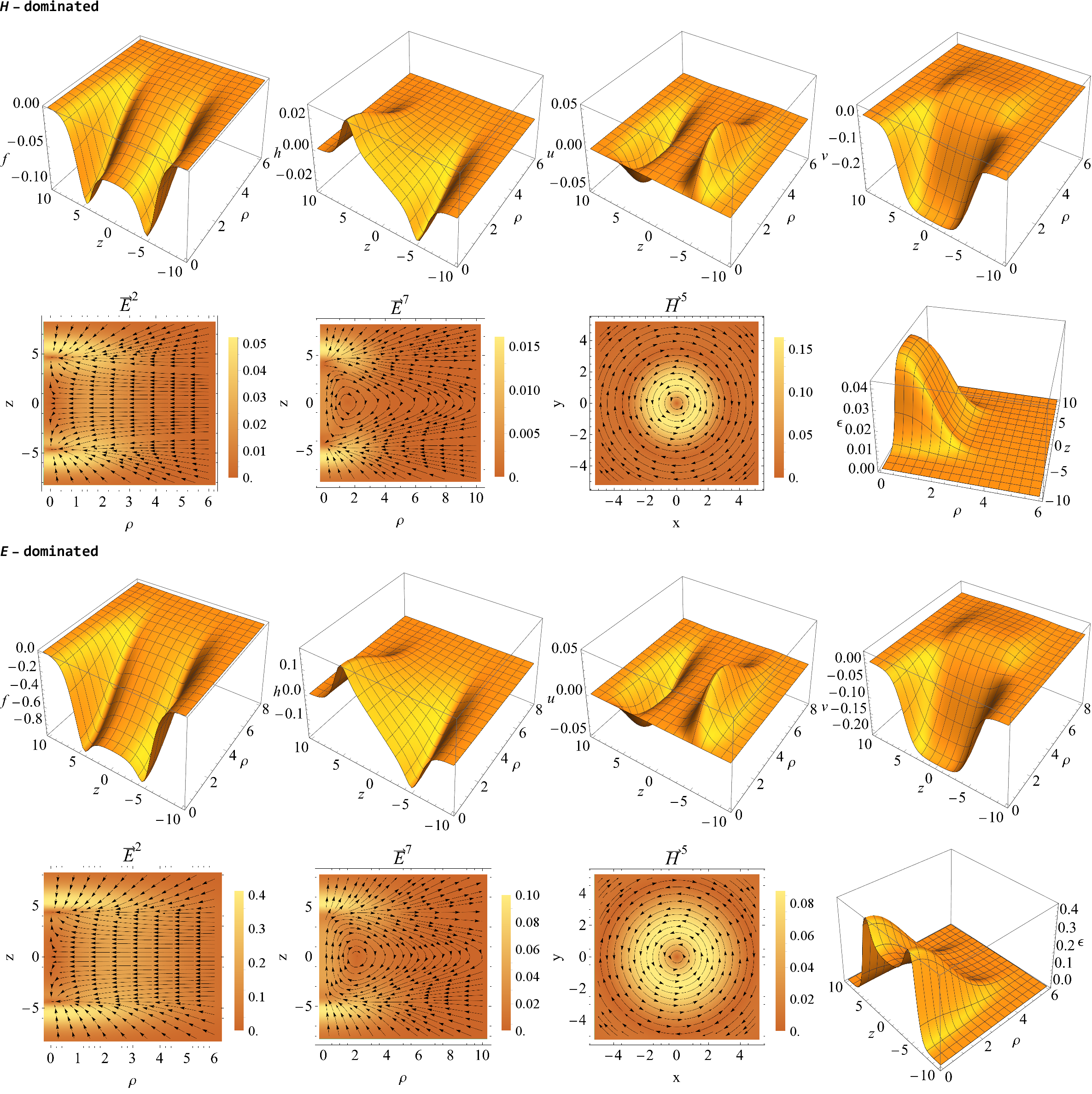}
	\end{center}
	\caption{Functions $f, h, u, v$, force lines for the electric $\vec{E}^2$, $\vec{E}^7$ and magnetic $\vec{H}^5$ fields [in Cartesian coordinates $(x,y)$], and the energy density $\varepsilon$ 
from Eq.~\eqref{energy_dens} for the magnetic-field-dominated system	
  (two top rows)  and for the electric-field-dominated system  (two bottom rows) with the choice $z_0=5$ (for the values of other parameters, see the caption of Fig.~\ref{fig_energy_flux}).
	}
\label{fig_sols_H_E_domin}
\end{figure}

\subsection{Asymptotic behavior}

Even before obtaining numerical solutions, one can estimate their asymptotic behavior, bearing in mind that all the 
functions $f, h, u,$ and $v$ are assumed to be  exponentially decaying with distance.
For this purpose, it is convenient to use a spherical coordinate system $\{r, \theta, \varphi\}$. In this case, as $r \rightarrow \infty$, 
 the functions $f, h, u, v\to 0$ exponentially fast. As a result, from Eqs.~\eqref{eqn_f}-\eqref{eqn_h}, one can find the following asymptotic equations:
\begin{align}
	\bigtriangleup_{r, \theta} (f,h,v) -	(f, h, v)= & 0 ,
\label{2_90}\\
  \bigtriangleup_{r, \theta} u - \frac{u}{r^2\sin^2{\theta}}  -	u =& 0 ,
\label{2_100}
\end{align}
where $\bigtriangleup_{r, \theta}$ is the Laplacian in the coordinates $r,\theta$. Eqs.~\eqref{2_90} for the functions $f,h,$ and $v$ have obvious solutions in the form
\begin{equation}
	(f, h, v) \approx  C_{(f,h,v)}
	\left( Y\right)^0_{l_{(f,h,v)}}
	\frac{e^{- r }}{r} ,
\label{2_120}
\end{equation}
where $\left( Y\right)^0_{l_{(f,h,v)}}$ are spherical functions and $C_{(f,h,v)}$ are constants. In turn, Eq.~\eqref{2_100} has a solution similar to \eqref{2_120}, 
but only with the angular part expressed in terms of special functions (we do not show this expression here to avoid overburdening the text). 
The general solution of Eqs.~\eqref{2_90} and \eqref{2_100} represents a superposition of the above solutions, and it is obtained by summing over $l_{(f,h,v)}$. 
Note that the results of numerical calculations obtained below indicate that the angular part $\left( Y\right)^0_{l_{(f,h,v)}} = \text{const}$. 

\subsection{Numerical results}

Depending on the choice of free parameters determining characteristics of the currents  \eqref{current_2t} and \eqref{current_5_rho}, 
one can obtain systems with different ratio of magnitudes of the electric and magnetic fields with different color indices. As an illustration, 
we consider  two systems where either the magnetic or electric field dominates. The corresponding numerical results of 
solving Eqs.~\eqref{eqn_f}-\eqref{eqn_h} are given in Fig.~\ref{fig_sols_H_E_domin} for the magnetic-field-dominated system (two top rows) 
and for the electric-field-dominated system (two bottom rows).

\begin{figure}[t]
	\begin{minipage}[t]{.49\linewidth}
		\begin{center}
			\includegraphics[width=.98\linewidth]{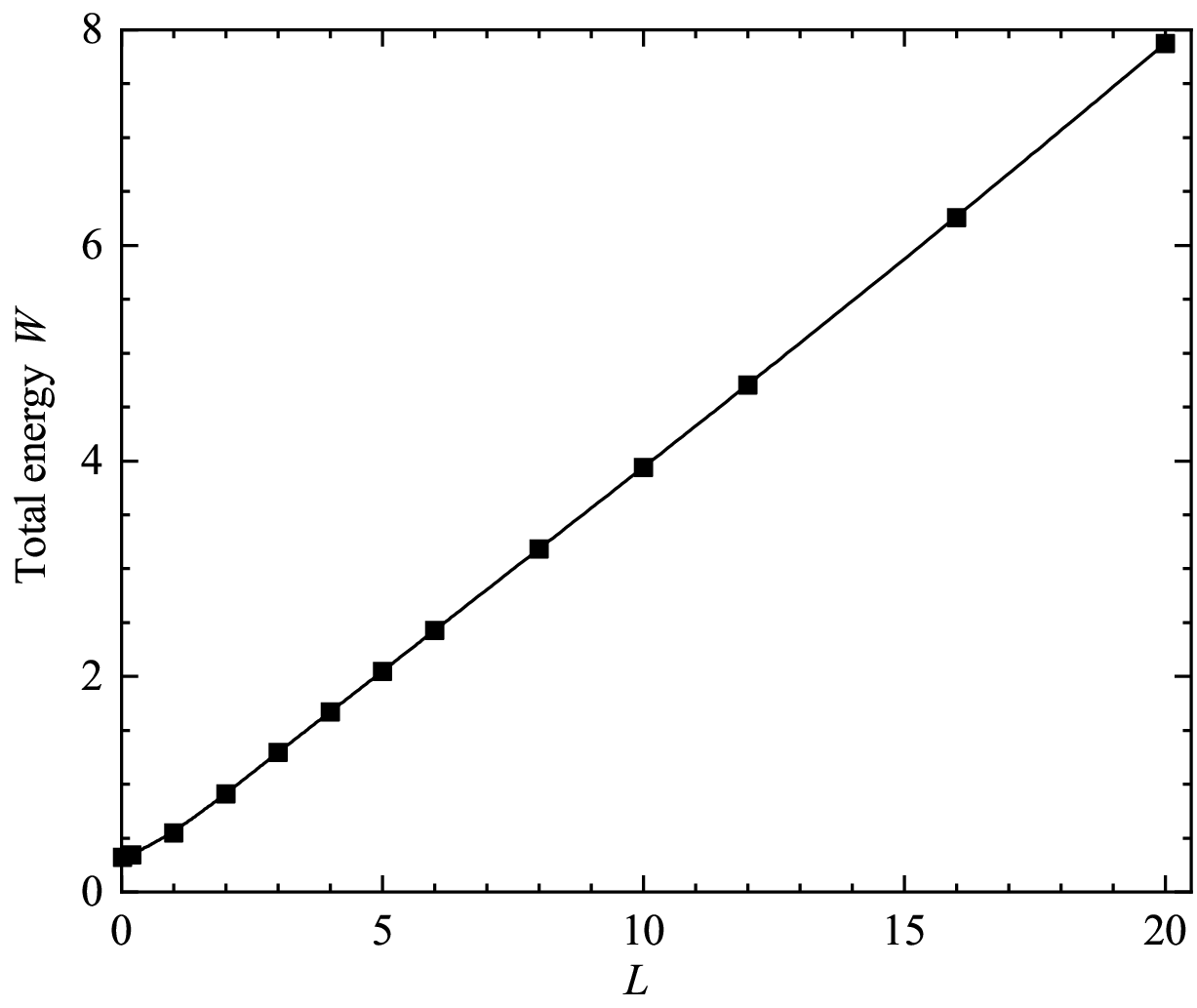}
		\end{center}
	\end{minipage}\hfill
	\begin{minipage}[t]{.49\linewidth}
		\begin{center}
			\includegraphics[width=1.\linewidth]{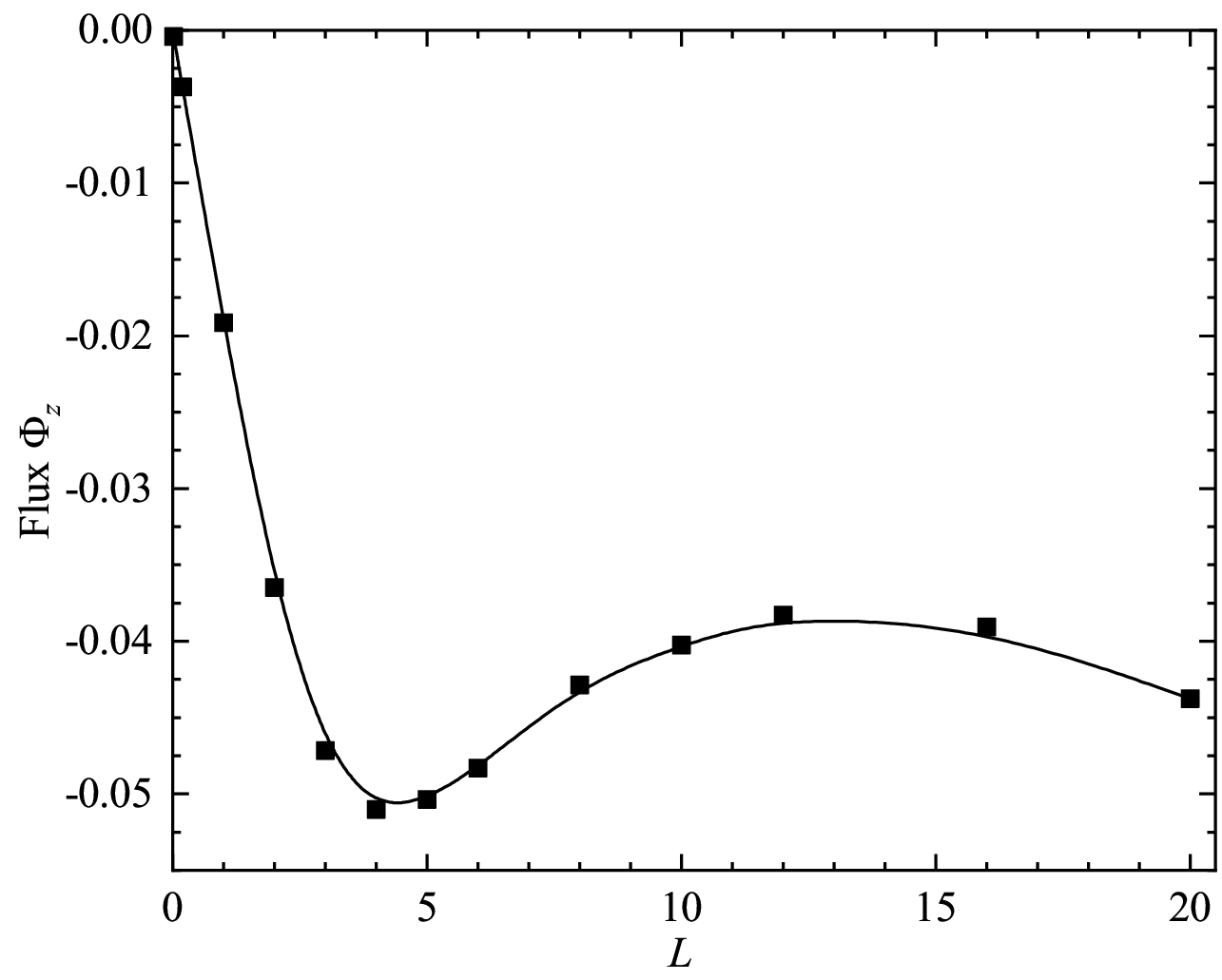}
		\end{center}
	\end{minipage}
	\begin{minipage}[t]{.49\linewidth}
		\begin{center}
			\includegraphics[width=.98\linewidth]{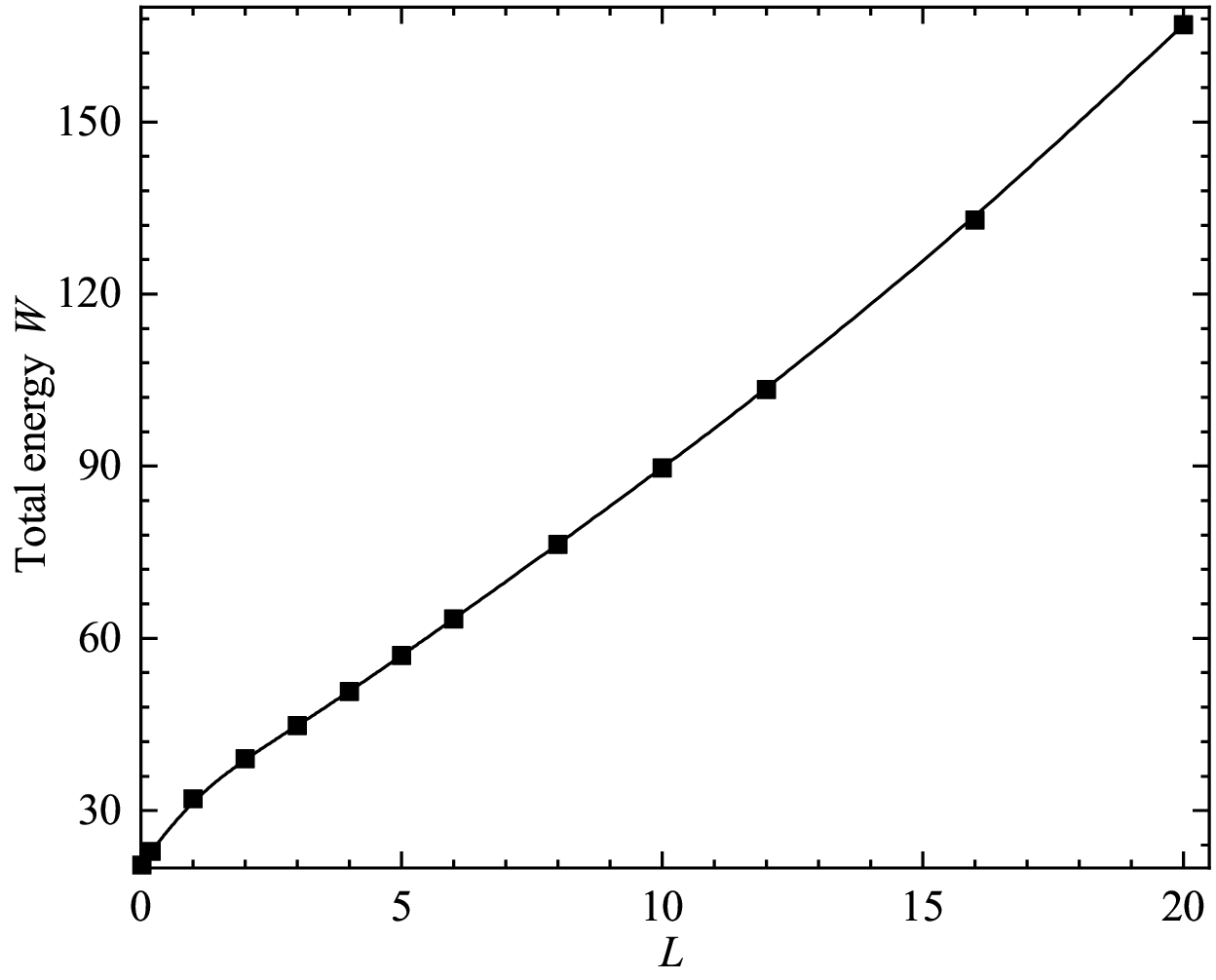}
		\end{center}
	\end{minipage}\hfill
	\begin{minipage}[t]{.49\linewidth}
		\begin{center}
			\includegraphics[width=1.\linewidth]{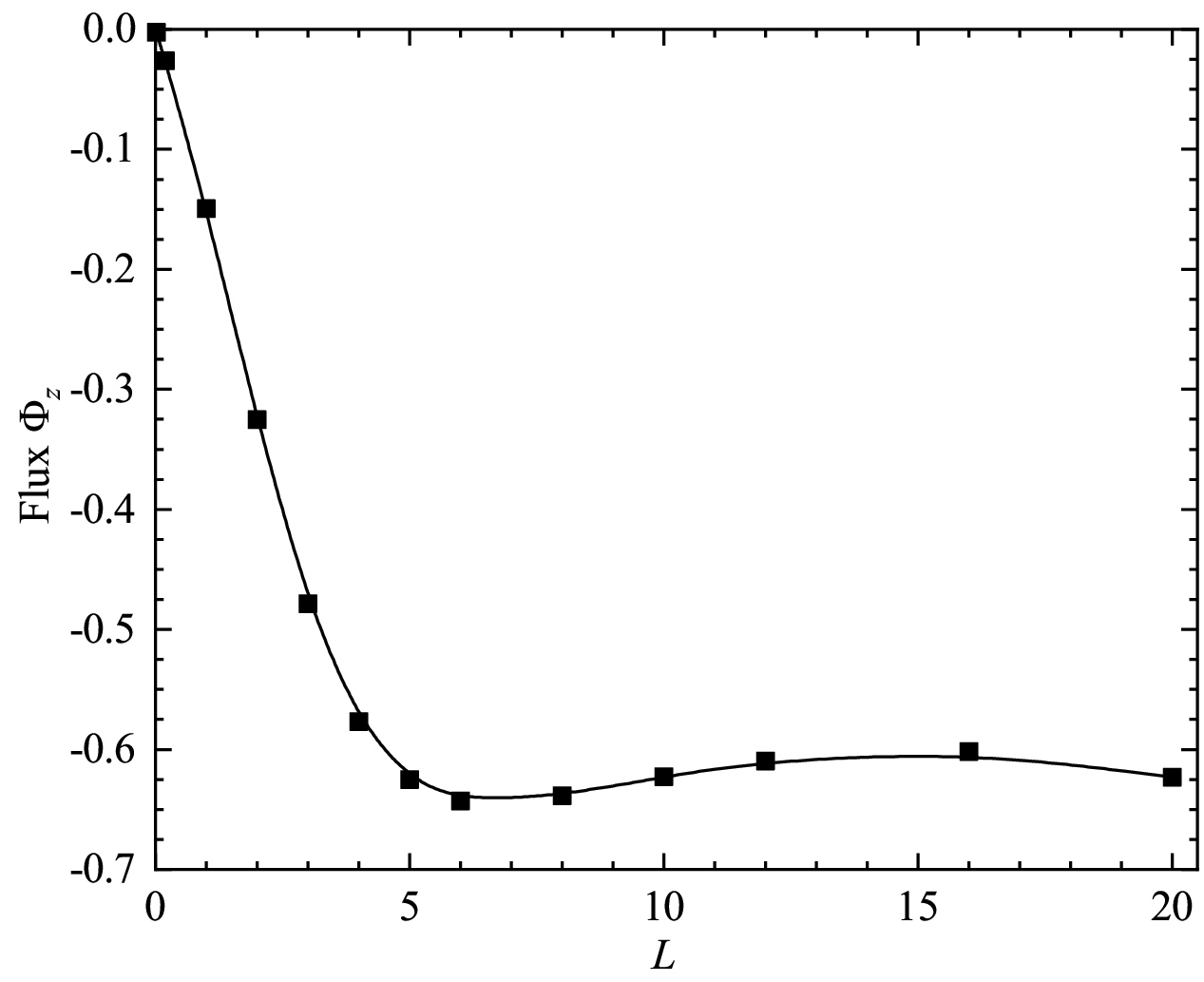}
		\end{center}
	\end{minipage}
	\caption{Total energy $W$ from Eq.~\eqref{total_energy}  and the flux of the longitudinal chromoelectric field $E^7_z$ from Eq.~\eqref{flux} as functions of $L=2 z_0$.
The top plots correspond to the magnetic-field-dominated systems with the choice $j_1=j_2=j_3=0.2, \rho_1=2, z_1=1$,  $j_4=-j_3$, and $j_{\text{glc}}=0.1$; 
the bottom plots correspond to the electric-field-dominated systems with the choice
 $j_1=j_2=1, j_3=0.1, \rho_1=3, z_1=1$,  $j_4=-j_3$, and $j_{\text{glc}}=1$.
	}
	\label{fig_energy_flux}
\end{figure}

For us, it is of especial interest to study the dependence of the total dimensionless energy of the systems under consideration on the distance between charges,
\begin{equation}
	W\equiv \frac{W_{\text{NU}}}{m} = 2 \pi \int \limits_{-\infty}^{\infty} dz \int \limits_{0}^{\infty} 
	\rho \varepsilon d \rho ,
\label{total_energy}
\end{equation}
[this expression is derived using Eq.~\eqref{energy_dens}, and $W_{\text{NU}}$ is a dimensional energy in natural units]
and the flux $\Phi_z$ of the color electric Proca field $\vec{E}^7$ through the plane $z = 0$,
\begin{equation}
	\Phi_z \equiv g \Phi_{z\text{NU}}= 2 \pi \int \limits_{0}^{\infty} \rho E^7_z d \rho 
\label{flux}
\end{equation}
[this expression is derived using Eq.~\eqref{fields_5} on account of the transition to the dimensionless electric field strength 
$E^7_z\to \left(m^2/g\right)E^7_z$ (for other components of the electric and magnetic field strengths the transition to the dimensionless variables is similar)]. 
The corresponding results of computations are given in Fig.~\ref{fig_energy_flux}.

\begin{figure}[t]
	\begin{center}
		\includegraphics[width=.5\linewidth]{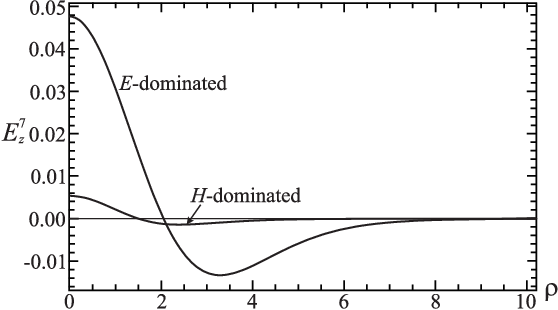}
	\end{center}
	\caption{The typical profiles of the longitudinal component  $E^7_z$ of the electric field in the plane $z = 0$ for the configurations with the magnetic field dominance  
 (shown in two top rows of Fig.~\ref{fig_sols_H_E_domin}) and with the electric field dominance  
(shown in two bottom rows of Fig.~\ref{fig_sols_H_E_domin}).
	}
\label{fig_E_7_z}
\end{figure}

It is seen from the results obtained that
\begin{itemize}
	\item The total energy $W$ increases linearly with increasing distance between the static charges; this is analogous to the behavior of the energy 
of interaction between two quarks in QCD when the distance between them increases. 
At the same time, the flux of the non-Abelian electric field $\vec{E}^7$ through the plane $z = 0$ first increases (modulus) with increasing distance between the charges,
and for large $L$ changes only slightly. 
	\item The behavior of the chromoelectric fields $\vec{E}^7$ and $\vec{E}^2$ is fundamentally different.  The field $\vec{E}^7$ creates a flux 
directed from one charge to another, and its  $z-$component has both the nonlinear, $f v/2$, and gradient, $ -h_{, z}$, terms directed in different directions 
and creating a total nonvanishing flux of this field through the plane $z = 0$. For the field $\vec{E}^2$, the situation differs in principle: 
the flux of this field through the plane $z = 0$ is equal to zero and both components (nonlinear, $-h v/2 $, and gradient,  $ - f_{, z}$) 
are equal to zero on the plane $z = 0$. Such a behavior is analogous to the behavior of force lines between charges with opposite signs in the first case and between
charges with the same sign in Maxwell's electrodynamics. 
	\item The electric fields $\vec{E}^{2, 7}$ and the magnetic field $\vec{H}^5$ have different color indices; this can be written as $E^a_i H^a_j = 0$ (no summation over $a$). 
	\item The profile of the longitudinal electric field $E^7_z$ obtained here has one node (see Fig.~\ref{fig_E_7_z}) and it
	differs from the nodeless profiles of the longitudinal electric field obtained using lattice calculations 
(see, e.g., Ref.~\cite{Baker:2021jnr} where the color electric  field is also sourced by static charges~-- quarks). 
One can naively expect that this difference may disappear for another choice of charge/current densities
 \eqref{current_2t}, \eqref{current_5_rho}, \eqref{current_7t}, and~\eqref{current_5z}. 
	\item The force lines of the color magnetic field create circles with a center coinciding with that of the flux tube. 
  \item The profile of force lines for the electric field  $\vec{E}^7 $ (see Fig.~\ref{fig_sols_H_E_domin}) shows that there are closed force lines, i.e. $\curl{\vec{E}^7} \neq 0$. 
  This is related to the fact that according to~\eqref{fields_5} $\vec{E}^7 = f \vec{A}^5/2 - \grad{h}$ and consequently 
    $ \curl{\vec{E}^7} = f \vec{H}^5/2 + \grad{f} \times \vec{A}^5$. Thus the presence of nonzero magnetic field
     $\vec{H}^5$ results in the appearance of nonzero  $\curl{\vec{E}^7} \neq 0$
   (the effect related to the nonlinearity of the non-Abelian field).
  \item It is seen from Fig.~\ref{fig_energy_flux} that the flux of the electric field through the plane $z=0$ changes slowly for quite large distances between the charges. 
  Apparently, this leads to the appearance of the practically linear dependence of the total energy on the distance. 
  \item The spatial distribution of the energy density of the systems with the electric field dominance differs considerably from that of the configurations with the magnetic field dominance 
  by the presence of two distinct maxima caused by a dominating contribution coming from the charge density compared with the current density.
\end{itemize}

It is interesting to note that, due to the nonlinearity of the system, even when the current $j^5$ is completely ``turned off'', there are configurations with the flux of the  longitudinal electric field which have the magnetic field that is much smaller than the electric field.

To conclude this section, let us estimate the magnitude of the Proca field mass $m$ that is needed to obtain a typical value of the tension coefficient of the string between quarks obtained in lattice calculations. Since we consider static charges, one can regard the total energy of the system $W$ calculated above as a corresponding interaction potential of the quarks that is calculated in lattice calculations in QCD. Then, for example, it is seen from the top left panel of Fig.~\ref{fig_energy_flux} that for the magnetic-field-dominated systems our potential increases linearly with distance between charges in the region
$L\gtrsim 2$: $W\sim K L$. On changing to dimensional units, this expression takes the form [see Eq.~\eqref{total_energy}]:
$$
W_{\text{NU}}=K_{\text{NU}} l ,
$$
where $l=L/m$ and $K_{\text{NU}}$ is a string tension coefficient (measured in natural units). It is known from lattice calculations that 
$K_{\text{NU}}\approx 0.2\, \text{GeV}^2$~\cite{Simonov:1996ati}. Then, to obtain this value of  $K_{\text{NU}}$ for the systems shown in the top left panel of Fig.~\ref{fig_energy_flux} 
(where $K\approx 0.38$), it is necessary to choose $m\approx 0.73\, \text{GeV}$. In turn, the dimensional distance between charges will then be determined by the expression
$$
l\approx 0.27\times 10^{-13}\cdot L\, \text{cm} .
$$
This quantity is in good agreement with a typical distance between quarks obtained in lattice calculations~\cite{Simonov:1996ati}.

\section{Discussion}
\label{discussion}

The results obtained tell us that in non-Abelian Yang-Mills-Proca theory with an external field source in the form of {\it ad hoc} currents there are  flux tubes filled 
with a longitudinal color electric field and having a total energy that increases linearly with increasing distance between static charges. 

Since similar behavior is inherent in a tube connecting quarks and threaded by a longitudinal electric field, it is of great interest to see the reasons of 
such a behavior of non-Abelian electric Proca fields. This would enable one to suggest some ideas concerning a possible relationship between these two theories 
(for instance, Proca theory could serve as an approximate description of QCD in the situation under consideration). These fields are sourced by the charge 
and current densities  \eqref{current_2t}, \eqref{current_5_rho}, \eqref{current_7t}, and \eqref{current_5z}. The charge density $j^2_t$ is concentrated 
at the points $z = \pm z_0$; this corresponds to the fact that at these points two charges are located. The static charge density $j^7_t$ is distributed over the whole space. 
The color current density $j^5_u$, as well as the charge density $j^2_t$, is concentrated at the points $z = \pm z_0$. In contrast, the current density  $j^5_z$ 
is distributed over the space. 

In such a situation, let us emphasize that there are charge/current densities concentrated near some center that corresponds to a 
location of a color charge/current; there are charge/current densities distributed over the space (this is especially evident for the current density $j^5_z$). This can be written in the form
 $$
	j^a_\mu = \left( j^a_\mu\right)_1 + \left( j^a_\mu\right)_2 . 
$$
Here the four-current $\left( j^a_\mu\right)_1$ is concentrated near the center of location of static charge/current (``quarks''), while the four-current $\left( j^a_\mu\right)_2$ is an extended object. 
Physically, this means that the four-current  $\left( j^a_\mu\right)_1$ is created by a charge/current localized near some center, while the current $\left( j^a_\mu\right)_2$ is created by 
some extended object which, as will be shown below, is related to a gluon condensate.

Our calculations indicate that in non-Abelian Proca theory there are flux tubes threaded by a longitudinal electric field and having a total energy that increases 
with increasing distance between static charges creating such a tube. Such result enables one to suppose that non-Abelian Proca theory under consideration 
with sources in the form of static charges is some approximation for QCD in describing a flux tube. In this case the four-current $\left( j^a_\mu\right)_1$ is created by quarks, 
while the four-current $\left( j^a_\mu\right)_2$ is an approximate description of a current created by a gluon condensate. 

The appearance of such a condensate may be 
approximately imagined as follows. In a quantum-averaged Lagrangian of  SU(3) Yang-Mills theory, there is the following term:
 \begin{equation}
	\left\langle 
		f^{abc} \hat{A}^b_\mu \hat{A}^c_\nu f^{amn} \hat{A}^{m \mu} \hat{A}^{n \nu} 
	\right\rangle .
\label{gluon_cond_1}
\end{equation}
Here $\left\langle \cdots \right\rangle $ denotes the quantum averaging; $\hat{A}^a_\mu (a, b, c = 2, 5 ,7)$ belongs to the sub-group SU(2): $\hat{A}^a_\mu \in \text{SU(2)} \subset \text{SU(3)}$ and 
$\hat{A}^m_\mu \in \text{SU(3)/SU(2)}$ with $m = 1, 3, 4, 6, 8$. The quantized gauge potentials 
$\hat{A}^a_\mu$ and $\hat{A}^m_\mu$ behave differently:
\begin{equation}
	\left\langle \hat{A}^a_\mu \right\rangle \approx A^a_\mu , 
	\left\langle \hat{A}^m_\mu \right\rangle = 0 
	\text{ but } 
	\left\langle \hat{A}^m_\mu \hat{A}^{m \mu} \right\rangle \neq 0 .
\label{quant_A}
\end{equation}
This enables one to say that the four-potential  $\hat{A}^a_\mu$ is almost classical, while $\hat{A}^m_\mu$ is purely quantum. 
In this case the expression~\eqref{gluon_cond_1} will approximately look like
$$
	\left\langle 
		f^{abc} \hat{A}^b_\mu \hat{A}^c_\nu f^{amn} \hat{A}^{m \mu} \hat{A}^{n \nu} 
	\right\rangle \approx f^{abc} A^b_\mu A^c_\nu 
		\left\langle 
		f^{amn} \hat{A}^{m \mu} \hat{A}^{n \nu} 
	\right\rangle . 
$$

Notice the following fact. On account of the antisymmetry of the structure constants $f^{a b c}$,  the expression $f^{amn} \hat{A}^{m \mu} \hat{A}^{n \nu} $ can be written in the form:
\begin{equation}
	f^{amn} \hat{A}^{m \mu} \hat{A}^{n \nu} = \frac{1}{2} f^{amn} 
	\left( 
	\hat{A}^{m \mu} \hat{A}^{n \nu} - \hat{A}^{n \mu} \hat{A}^{m \nu}
	\right) =  \frac{1}{2} f^{amn} \left[ \hat{A}^{m \mu}, \hat{A}^{n \nu}\right]_{m,n} ,
\label{m_n_commutator}
\end{equation}
where $[ \cdots ]_{m,n}$ is a commutator for indices $m, n$. For free noninteracting fields similar commutator
$$
	\left[ \hat{A}^{m \mu}(t, x^i), \hat{A}^{n \nu}(t, y^j)\right] = 0 
$$
for the points $x^i, y^j$ separated by a spacelike interval. For strongly interacting fields the commutator \eqref{m_n_commutator} can be nonzero, since for  them there can exist a static regular distribution of the field in space. This means that
$$
	\left[ \hat{A}^{m}_\mu(t, x^i), \hat{A}^n_\nu (t, y^j)\right]_{m,n} = 
	D^{m n}_{\mu \nu} \left( x^i, y^j\right) \neq 0 .
$$
Thus in a quantum-averaged Lagrangian of  SU(3) Yang-Mills theory there appears the following term:
$$
	f^{abc} f^{amn} A^b_\mu A^c_\nu D^{m n}_{\mu \nu} \left( x^i, x^j\right) . 
$$
In our opinion this term may describe a gluon condensate created by the quantum components  $\hat{A}^m_\mu$ of a non-Abelian gauge field. Note also that if the asymptotic behavior of the commutator is 
\begin{equation}
	D^{m n}_{\mu \nu} \left( 
	x^i \rightarrow \infty, x^j \rightarrow \infty 
	\right) \rightarrow m^2 , 
\label{Proca_mass}
\end{equation}
then \textit{the Proca mass $m$ is a purely quantum phenomenon appearing in a strongly nonlinear regime of QCD in nonperturbative quantization (the nonperturbative generation of the Proca mass)}. In this connection, note that Ref.~\cite{Hell:2021oea} discusses the question of  the massless limit in Proca theory, and it is shown that the apparent discontinuity in the massless limit is only an artefact of the perturbation theory.

The above discussion justifies the assumption that the four-current $\left( j^a_\mu\right)_1$ is created by static quarks, \textit{while the current $\left( j^a_\mu\right)_2$ is created by a gluon condesate}, 
i.e., it is a purely quantum object created by the quantum components $\hat{A}^m_\mu$. Note that this assumption, as well as the assumption \eqref{quant_A} concerning the different behavior of 
different components of the gauge potential $A^B_\mu$ (with $B = 1, \ldots , 8$) and the assumption \eqref{Proca_mass} concerning the nature of the Proca mass $m$, can be verified using lattice calculations.

A very important and interesting question is that of gauge invariance in the case under consideration.
The results concerning the flux tube were obtained in non-Abelian  SU(2) Yang-Mills-Proca theory that is clearly not a gauge-invariant theory.
In this section we consider an assumption that the present Proca theory is some approximation for a real quantum SU(3) theory. 
In such a case the divergent equation~\eqref{1_120} can be thought of as a choice of the gauge in initial  SU(3) theory where a contribution coming from the gluon condensate 
created by the fields $\hat{A}^m_{\mu} \in \text{SU(3)/SU(2)}$ is taken into account. This enables one to assume that for such a nonpertubative quantization the gauge invariance is conserved, but,
of course,  this assumption requires a more careful analysis. 

The discussion presented in this section enables one to assume that \textit{an approximate description of a flux tube can be made using classical equations within Yang-Mills-Proca theory where one takes into account the fact that these fields are sourced not only by color charges but also by a gluon condensate of quantum fields with zero average value and with nonzero dispersion. }

\section{Comparison with the results of lattice calculations
}
\label{comparison}

Using lattice calculations, it was shown in Ref.~\cite{Baker:2019gsi} that the simulated chromoelectric  
field $\vec{E}$ consists of the ``nonperturbative'',  $\vec{E}^{\text{NP}}$, and ``Coulomb'', $\vec{E}^\text{C}$, fields:
$$
\vec{E} = \vec{E}^{\text{NP}} + \vec{E}^{\text{C}} . 
$$

In order to compare this expression with the expressions for electric and magnetic fields, let us rewrite the expressions~\eqref{fields_5} 
and \eqref{fields_10} for these fields on account of a possible relationship between quantum SU(3) gauge theory and its approximate description in the form of SU(2)
Proca theory discussed in Sec.\ref{discussion}: 
\begin{align}
	\vec{E}^2 = & - \frac{h \vec{A}^5}{2} - \frac{\grad{f}}{g} + g f_{2 m n} 
	\left\langle 
		\hat{A}^m_{t} \hat{\vec{A}}^n 
	\right\rangle = \left( \vec{E}^2 \right)_{\text{nl}}
	+ \left( \vec{E}^2 \right)_{\grad} 
	+ \left( \vec{E}^2 \right)_{\text{gc}} \quad \text{with} \quad
	 m, n = 1, 3, 4, 6 ; 
\label{el_field_gl_cond_1}\\
	\vec{E}^7 = & \frac{f \vec{A}^5}{2} - \frac{\grad{h}}{g} + g f_{7 m n} 
	\left\langle 
		\hat{A}^m_{t} \hat{A}^n_{z} 
	\right\rangle = \left( \vec{E}^7 \right)_{\text{nl}}
	+ \left( \vec{E}^7 \right)_{\grad} 
	+ \left( \vec{E}^7 \right)_{\text{gc}} \quad \text{with} \quad m, n = 1, 3, 4, 6, 8 ; 
\label{el_field_gl_cond_2}\\
	H^5_\varphi = & \rho 
	\frac{ \left(v_{, \rho} - u_{, z}\right)}{g} + g \rho f_{5 m n} 
	\left\langle 
		\hat{A}^m_{\rho} \hat{A}^n_{z} 
	\right\rangle = \left( H^5_\varphi\right)_{\grad} 
	+ \left(H^5_\varphi \right)_{\text{gc}} \quad \text{with} \quad m, n = 1, 3, 4, 6, 8 . 
\label{el_field_gl_cond_5}
\end{align}
Here the first terms on the right-hand sides of Eqs.~\eqref{el_field_gl_cond_1} and \eqref{el_field_gl_cond_2} are nonlinear terms, the second terms are gradient terms, and the third terms are
associated with the presence of a gluon condensate [recall that the indices $2,5,7 \in \text{SU(2)} \subset \text{SU(3)}$ and $m,n \in \text{SU(3)/SU(2)}$.  Similarly, for the magnetic field~\eqref{el_field_gl_cond_5}].

It is seen that the ``nonperturbative'' component $\vec{E}^{\text{NP}}$ from lattice calculations coincides with the nonlinear plus gluon terms of the total vector of the chromoelectric field $\vec{E}^7$, 
$
	\left( \vec{E}^{ 7} \right)_{\text{nl}} +  \left( \vec{E}^{ 7} \right)_{\text{gc}} 
$, 
and correspondingly the ``Coulomb'' field $\vec{E}^{\text{C}}$ coincides with the gradient term $\left( \vec{E}^7 \right)_{\grad}$. 

In Ref.~\cite{Baker:2019gsi}, it was also shown that the magnetic field obtained using lattice calculations is negligibly small compared with the electric field. 
It must be mentioned that in those calculations the comparison was made for magnetic and electric fields possessing the same color index. In our calculations 
we deal with the similar situation: there are nonvanishing color electric fields $\vec{E}^{2, 7}$ and color magnetic fields having the same color index 
but equal to zero, $\vec{H}^{2, 7} = 0$. On the other hand, in our calculations there is nonzero magnetic field  $\vec{H}^{5} \neq 0$ and there is no a corresponding electric field: $\vec{E}^{5} = 0$. 
This situation can be described as $E^a_i  H^a_j = 0$ (no summation over $a$).
Notice that this result may be verified using lattice calculations. 

As noted in the previous section, in lattice calculations the profile of the longitudinal chromoelectric field differs from that of obtained by us in that in our case there is one node. This is related to the fact that there is a nonzero curl of the field 
$\vec{E}^7 = \left( \vec{E}^7 \right)_{\text{nl}} + \left( \vec{E}^7 \right)_{\grad}$ used in our calculations: 
$\curl{\vec{E}^7} = f \vec{H}^5/2 + \grad{f} \times \vec{A}^5$. Due to the presence of the quite large chromomagnetic field $ \vec{H}^5$ there appear closed force lines of the field $\vec{E}^7$; 
this leads to the appearance both of positive and of negative values of $E^7_z$. In turn, in lattice calculations this magnetic field is small, and therefore there are no closed force lines of the field~$\vec{E}^7$. 
Naively we may assume that this situation with one node can be changed by choosing profiles of charge/current densities different from those of given by Eqs.~\eqref{current_2t}, \eqref{current_5_rho}, \eqref{current_7t}, and \eqref{current_5z}.

\section{Conclusions}
\label{concl}

Summarizing the results obtained,
\begin{itemize}
	\item Within non-Abelian Yang-Mills-Proca theory, we have shown that there are regular solutions supported by static charges and currents. These solutions describe a flux tube threaded by a color longitudinal electric field with a color index  $a = 2, 7$, and force lines of a color magnetic field with another color index $a = 5$ are wound on the tube. 
	\item These flux tubes have a linear dependence of their total energy on the distance between static charges; this enables us to say about the effect similar to the confinement phenomenon in QCD.
	\item The above dependence of the total energy on the distance between charges is a consequence of the presence of the Proca mass which ensures an exponential decrease of the fields at infinity, as well as because of the nonlinear interaction between fields inherent in Yang-Mills theory.   
	\item The flux of the electric field through the plane $z = 0$ depends strongly on the length of the tube: when the length increases, the flux becomes a slowly varying quantity. 
  \item Our calculations indicate that there is a longitudinal chromoelectric field that ensures the confinement  of the charges.
   \item It is shown that the results obtained are in good agreement with the results of lattice calculations. 
   \item The question of a possible relationship between non-Abelian Yang-Mills-Proca theory employed here and quantum chromodynamics is considered. 
\end{itemize}

As a result, we have obtained the distribution of chromoelectric and chromomagnetic fields bearing a fairly close resemblance to the properties of the same fields that are needed for the confinement of quarks in QCD. 
We have discussed this interesting fact in Sec.~\ref{discussion}, where a possible scenario for a situation when non-Abelian Proca theory may serve as an approximate description of QCD has been considered. 

In future studies it would be interesting, by comparing with lattice calculations, to obtain more realistic profiles of the currents which could provide better agreement of our model with lattice calculations. 
For example, one might verify the existence  of a chromomagnetic field $\vec{H}^5$ which is perpendicular to the chromoelectric field
 $\vec{E}^{2,7}$	($\vec{H}^5 \perp \vec{E}^{2,7}$) in color space in the sense that  $H^a_i E^a_j = 0$ (no summation over $a$).



\end{document}